\documentclass[twocolumn,amsmath,amssymb,aps]{revtex4}
\usepackage{graphicx}
\usepackage{amsmath,amssymb}
\usepackage{mathrsfs}
\usepackage{color}
\usepackage{afterpage}
\usepackage[version=3]{mhchem}
\usepackage{natbib}
\usepackage{soul}
\usepackage[caption=false]{subfig}
\usepackage{array}
\usepackage{multirow}
\usepackage[colorlinks, citecolor=blue,urlcolor=blue, linkcolor=blue, bookmarks=false]{hyperref}
\hypersetup{colorlinks=true , citecolor=blue, urlcolor=blue, linkcolor=blue}
\usepackage{braket}
\usepackage{bm}
\begin{document}

\title{Probing switchable valley-related Hall effects in 2D magnetic MXenes}

\author{Ankita Phutela\footnote{ankita@physics.iitd.ac.in}, Sajjan Sheoran, Saswata Bhattacharya\footnote{saswata@physics.iitd.ac.in}} 
\affiliation{Department of Physics, Indian Institute of Technology Delhi, New Delhi 110016, India}
\begin{abstract}
	\noindent 

The search for two-dimensional materials with exotic valley-dependent properties has attracted rapid attention as they are fundamentally intriguing and practically appealing for nanoscale device applications. Here, using first-principles calculations, we report the identification of promising intrinsic valley-related switchable Hall effects in Cr$_2$CSF. With a high out-of-plane magnetic anisotropy, Cr$_2$CSF is a ferrovalley semiconductor with spontaneously polarized valleys having a valley polarization of 27.1 meV in the conduction band. This facilitates the observation of an intrinsic anomalous valley Hall (AVH) effect that is manipulated under an in-plane electric field. The underlying physics of spontaneous valley polarization is also discussed based on the SOC Hamiltonian model. Furthermore, on application of unidirectional compressive strain, Cr$_2$CSF is further transitioned from the AVH phase to a long-sought quasi-half valley metal state. Here, only the electrons are valley polarized such that holes and electrons carriers are separated under the in-plane electric field. Our work enriches materials with the valley-related Hall effects and provides a platform for interplay among valleytronics and spintronics.



 
\end{abstract}
\maketitle

\section{Introduction}
Valley, the energy extreme of valence or conduction bands, in a two-dimensional (2D) lattice can be used as a degree of freedom for the carriers in condensed matter materials ~\cite{rycerz2007valley,shkolnikov2002valley}. Given the considerable separation in the momentum space, scattering between these valleys can be effectively suppressed, allowing electrons, holes, or excitons to be manipulated for a longer period of time. Therefore, the valley index is robust against the long-range impurity scattering ~\cite{xiao2007valley,schaibley2016valleytronics,gunawan2006valley}. Among 2D materials, the transition metal dichalcogenides have acquired significant attention due to their exceptional and tunable properties ~\cite{sheoran2023probing,sheoran2023coupled,srivastava2015valley,yao2008valley,macneill2015breaking,zhao2017enhanced,zeng2018exploring}.
In these strong spin-orbital coupling (SOC) systems, the valley degree of freedom is present due to the inversion asymmetry. To make use of the valley index of these nonmagnetic materials as an information carrier, it is essential to remove the energy degeneracy between inequivalent valleys by breaking the time reversal symmetry ~\cite{guo2023possible}. Subsequently, attempts have been made to extrinsically induce valley polarization by employing various approaches, such as optical pumping ~\cite{srivastava2015valley,yao2008valley}, application of magnetic fields ~\cite{macneill2015breaking}, utilization of magnetic substrates ~\cite{zhao2017enhanced}, and introduction of magnetic doping ~\cite{zeng2018exploring}, have been employed. Nevertheless, these methods are quite fragile and often come at the cost of altering the intrinsic energy band structures and crystal configurations.

Therefore, to conquer these challenges, it is urgent to seek materials showing anomalous valley Hall (AVH) effect rooted in spontaneous valley polarization. Indeed, there has been an exciting development in this field in the form of ferrovalley (FV) semiconductors, which break both time-reversal and space-inversion symmetries. These FV materials exhibit this unique characteristic, offering an unprecedented solution to the aforementioned challenges and hold significant potential for advancements in valleytronics applications. However, such materials are currently scarce and only a handful of potential systems have been suggested which include 2H-VSe$_2$ ~\cite{tong2016concepts}, 2H-FeCl$_2$  ~\cite{zhao2022intrinsic}, Nb$_3$I$_8$ ~\cite{peng2020intrinsic}, VSi$_2$N$_4$ ~\cite{cui2021spin}, and GdI$_2$ ~\cite{cheng2021two}. Nevertheless, most of these existing systems suffer from in-plane magnetization ~\cite{tong2016concepts,peng2020intrinsic,cheng2021two}, which substantially hinders the emergence of natural valley polarization. Therefore, there exists a pressing need to explore new candidates that exhibit spontaneous valley polarization, particularly those inherently characterized by robust out-of-plane magnetization.\\
\indent MXenes, which are 2D transition metal carbon and nitrogen compounds, have gained much attention following their successful experimental preparation ~\cite{naguib2011two,naguib201425th,naguib2021ten}. 
The general formula of MXene can be written as M$_{n+1}$X$_n$T $(n = 1-3)$, where M denotes a transition metal; X denotes C or N; T denotes surface terminations like Cl, F, O, OH ~\cite{naguib2012two}. They have been investigated in applications like nanocomposites and hybrid materials ~\cite{firouzjaei2022mxenes,jiang2022improving}. More recently, they have been proposed in valleytronics and only a few systems have been studied for their valley-dependent properties ~\cite{lu2022enhanced,feng2022valley}. Ferromagnetic (FM) MXenes like Cr$_2$C display an adjustable magnetism upon symmetric or asymmetric surface functionalization along with high Curie or N\'eel temperatures ~\cite{li2021intrinsic}.\\
\indent In this study, we present the discovery of various intrinsic valley-related Hall effects in 2D Cr$_2$CSF MXene based on first-principles calculations. Our findings reveal that this material acts as a FM semiconductor, with band edges situated at the $+$K and $-$K points. Of paramount importance is the spontaneous polarization of these valleys and the presence of substantially high out-of-plane magnetization which is also shown by SOC model Hamiltonian studies. Moreover, uniaxial strain yields quasi-half valley metal (QHVM) phase in Cr$_2$CSF monolayer.

\section{Computational Methods}
The calculations are performed at density functional theory (DFT)~\cite{hohenberg1964inhomogeneous,kohn1965self} level with the projector augmented wave (PAW)~\cite{kresse1999ultrasoft,blochl1994projector} method implemented in Vienna \textit{ab initio} Simulation Package (VASP)~\cite{kresse1996efficient} code.
The exchange-correlation interactions are treated by generalized gradient approximation (GGA) in the form of the Perdew-Burke-Ernzerhof (PBE) functional ~\cite{perdew1996generalized}. The cutoff energy of 520 eV is used for the plane wave basis set such that the total energy calculations are converged within 10$^{-5}$ eV. The vacuum in \textit{z}-direction is set to 30 \AA\ to avoid artificial interactions caused by the periodic boundary conditions. The 11$\times$11$\times$1 \textit{k}-grid is used to sample the 2D Brillouin zone. All the structural parameters are fully optimized until the Hellmann-Feynman forces are smaller than 1 meV/\AA. Considering the strong correction effect for the localized 3\textit{d} electrons of Cr atoms, the GGA+U method is employed ~\cite{rohrbach2003electronic}. This is implemented using Dudarev et al.'s rotationally invariant approach where only the effective U (U$_{eff}$) based on the difference between the on-site Coulomb interaction parameter and exchange parameters is meaningful ~\cite{dudarev1998electron}. Following the previous works ~\cite{wang2006oxidation,he2021two}, the U value for Cr 3\textit{d} orbitals is chosen as 3.5 eV. Hybrid functional Heyd-Scuseria Ernzerhof (HSE06) is also used to confirm the results of GGA+U calculations ~\cite{heyd2003hybrid}. The dynamical stabilities are calculated using density functional perturbation theory using PHONOPY code ~\cite{baroni2001phonons} with a  6$\times$6$\times$1 supercell. Berry curvatures are obtained from the maximally localized Wannier functions as implemented in WANNIER90 ~\cite{mostofi2008wannier90}. 




\section{RESULTS AND DISCUSSION}
\begin{figure}[htp]
	\includegraphics[width=0.5\textwidth]{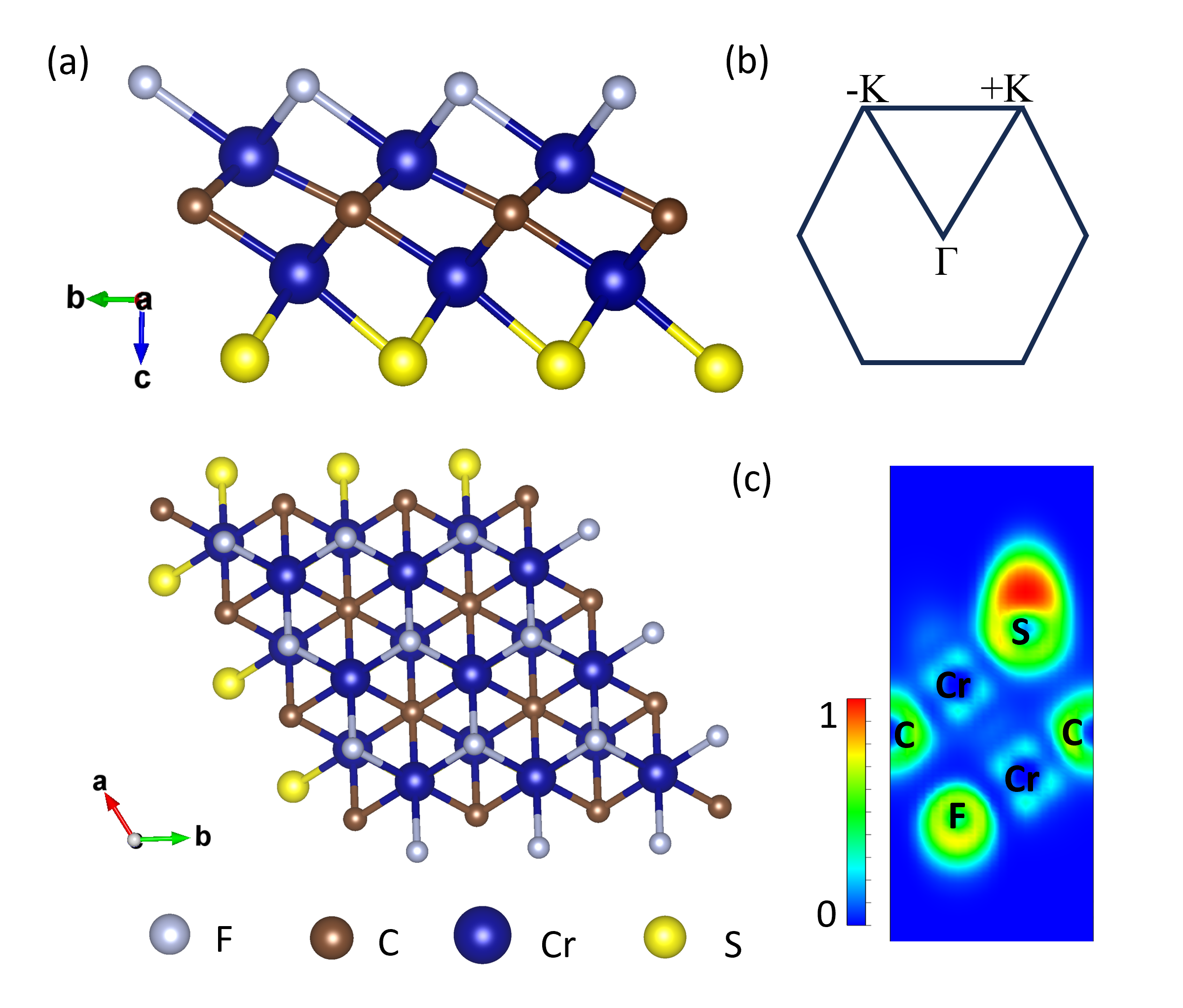}
	\caption{(a) Side and top views of Cr$_2$CSF. (b) Hexagonal 2D Brillouin zone. (c) Electron localization function map of Cr$_2$CSF, which is renormalized to values between 0 for charge depletion and 1 for charge accumulation.}
	\label{pic1}
\end{figure}

Cr$_2$CSX (X = F, O, H, Cl, Br, and I) can be surface engineered by functionalizing Cr$_2$C MXene in different ways, generating four distinct structures in each case [see Sec. I of Supplemental Material (SM)] ~\cite{li2021intrinsic}. Type II is the most energetically stable configuration among all Janus Cr$_2$CSX MXenes [see Sec. II of SM].
We have taken Cr$_2$CSF as a prototype for studying the valley properties because of its higher value of spontaneous valley polarization and other promising characteristics which will be discussed in the subsequent sections. The energetic stability of Type II structure of Cr$_2$CSF has also been confirmed by HSE06 functional calculations [see Sec. III of SM]. Cr$_2$CSF exhibits a hexagonal lattice structure characterized by the \textit{P}3\textit{m}1 space group [Fig. \ref{pic1}(b)]. Each unit cell contains one C, one S, one F, and two Cr atoms, which are stacked in the sequence of F-Cr-C-Cr-S. To probe the bonding nature of Cr$_2$CSF MXene, we have calculated the electron localization function within the plane encompassing F-Cr-C-Cr-S. Fig. \ref{pic1}(c) suggests an ionic bonding character for all the bonds as the electrons are predominantly localized around the atoms. 

\begin{figure}[htp]
	\includegraphics[width=0.5\textwidth]{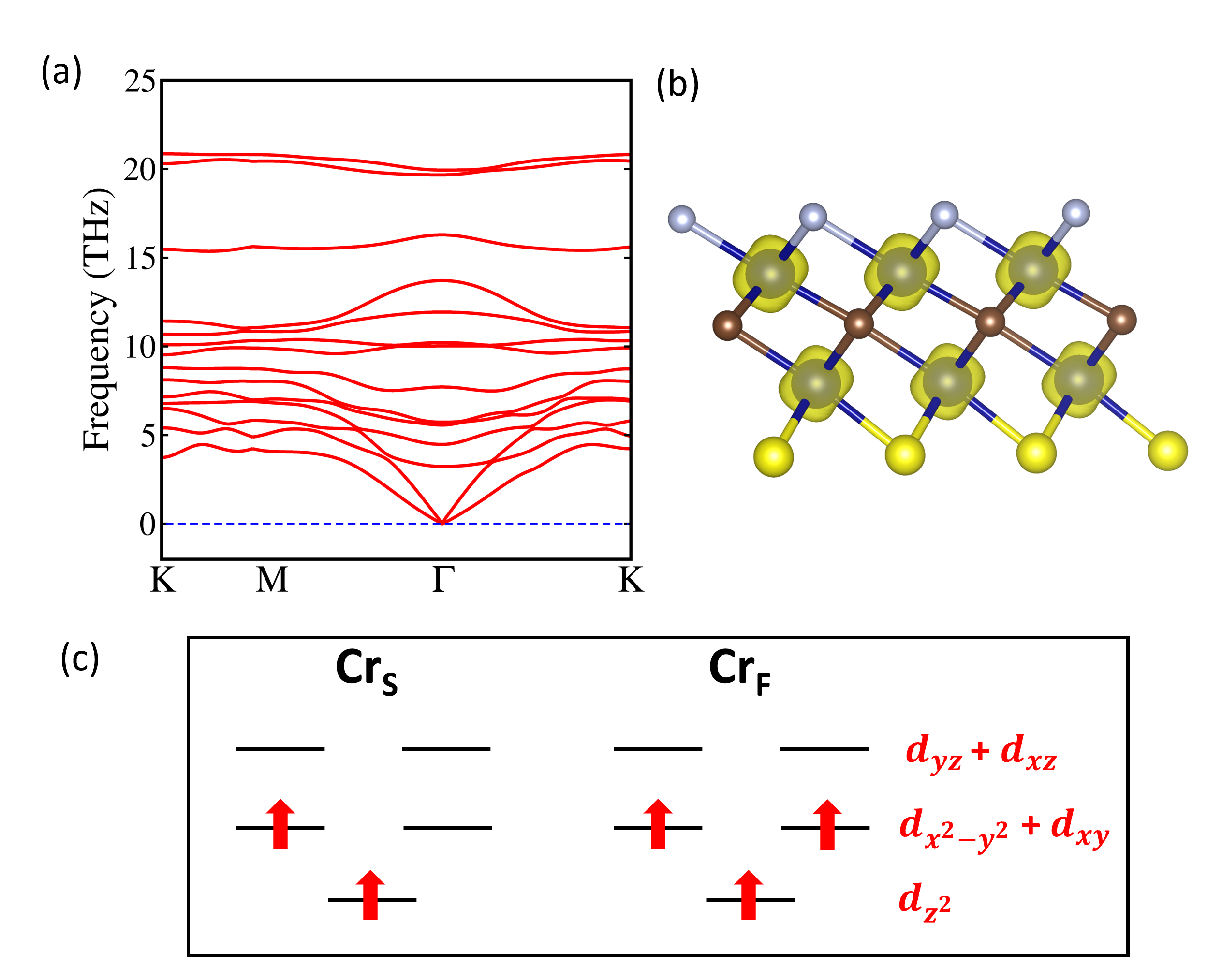}
	\caption{(a) Phonon dispersion map of Cr$_2$CSF. (b) Spin-polarized charge densities on Cr$_2$CSF MXene, where spin-up density is shown in yellow. (c) The predicted occupations of \textit{d} orbitals on Cr$_\textup{S}$ and Cr$_\textup{F}$ atoms. }
	\label{pic2}
\end{figure}

Afterwards, the absence of negative frequencies in the phonon dispersion plot of Cr$_2$CSF confirms the dynamical stability [Fig. \ref{pic2}(a)]. The spin-polarized charge density for Cr$_2$CSF is illustrated in Fig. \ref{pic2}(b), depicting the spontaneous spin polarization. Further, to ascertain the preferred magnetic ground state configuration of Cr$_2$CSF MXene, we have analyzed the FM and three antiferromagnetic (AFM) states [see Sec. IV of SM]. Cr$_2$CSF prefers the FM coupling over AFM couplings. The valence electronic configuration of the Cr atom is 3d$^5$4s$^1$. According to the local symmetry of the crystal field surrounding the Cr magnetic ions, the five fold degenerate 3\textit{d} orbitals split into three categories i.e., $A^\prime$(\(d_{z^2}\)), $E^\prime$(\(d_{yz} + d_{xz}\)) and $E^{\prime\prime}$(\(d_{x^2 - y^2} + d_{xy}\)), as shown in Fig. \ref{pic2}(c) ~\cite{he2016new}. Upon functionalization, both the Cr atoms are coordinated to different atoms. Cr$_\textup{F}$ atoms (Cr atom bound to F atom) have the electronic configuration of \(3{d^3}4{s^0}\) while the electronic configuration of Cr$_\textup{S}$ atoms (Cr atom bound to S atom) is \(3{d^2}4{s^0}\). As anticipated, the calculated magnetic moment per unit cell for Cr$_2$CSF is 5 \( \mu_B \) where \( \mu_B \) (Bohr magneton) where Cr$_\textup{F}$ and Cr$_\textup{S}$ atoms have the magnetic moment 2.9 \( \mu_B \) and 2.7 \( \mu_B \), respectively. This preferred FM coupling in Cr$_2$CSF is attributed to the double-exchange mechanism as the magnetic exchange arises between the ions in different oxidation states ~\cite{li2021intrinsic}.

Furthermore, the magnetic anisotropy energy (MAE) of Cr$_2$CSF MXene has been calculated to determine its magnetization easy axis. MAE is defined as the energy difference between the in-plane and the out-of-plane magnetization: MAE = E$_{x/y}$ $-$ E$_z$. The negative/positive MAE means an easy axis along the in-plane/out-of-plane direction. Due to C$_{3v}$ point group symmetry, energy is degenerate for magnetization along arbitrary in-plane direction. Therefore, we have chosen in-plane direction to be along \textit{x} direction ~\cite{sheoran2024multiple}. The MAE for monolayer Cr$_2$CSF is calculated to be 1.16 meV per unit cell, suggesting that its easy axis is along the out-of-plane direction. This substantially high MAE value suggests that the out-of-plane magnetization is robust. This stability is advantageous for establishing the long-range magnetic order, and is tantalizing for realizing the spontaneous valley polarization that has been discussed later. The observed value of MAE in our material exceeds that of Ni, Fe, and Co, which typically have MAE values of the order 1 \(\mu\)eV per unit cell ~\cite{halilov1998magnetocrystalline} and higher than that for VSi$_2$P$_4$ ~\cite{feng2021valley}, Cr$_2$Se$_3$ ~\cite{he2021two}, and Cr$_2$S$_3$  ~\cite{li2023spontaneous}. Evidently, in materials suffering from in-plane magnetization, achieving out-of-plane magnetization through external methods is challenging ~\cite{qi2015giant,zhao2017enhanced,zhong2017van}.  However, in Cr$_2$CSF no such external means are required, making it more superior for practical applications. 
 
\begin{figure}[htp]
	\includegraphics[width=0.5\textwidth]{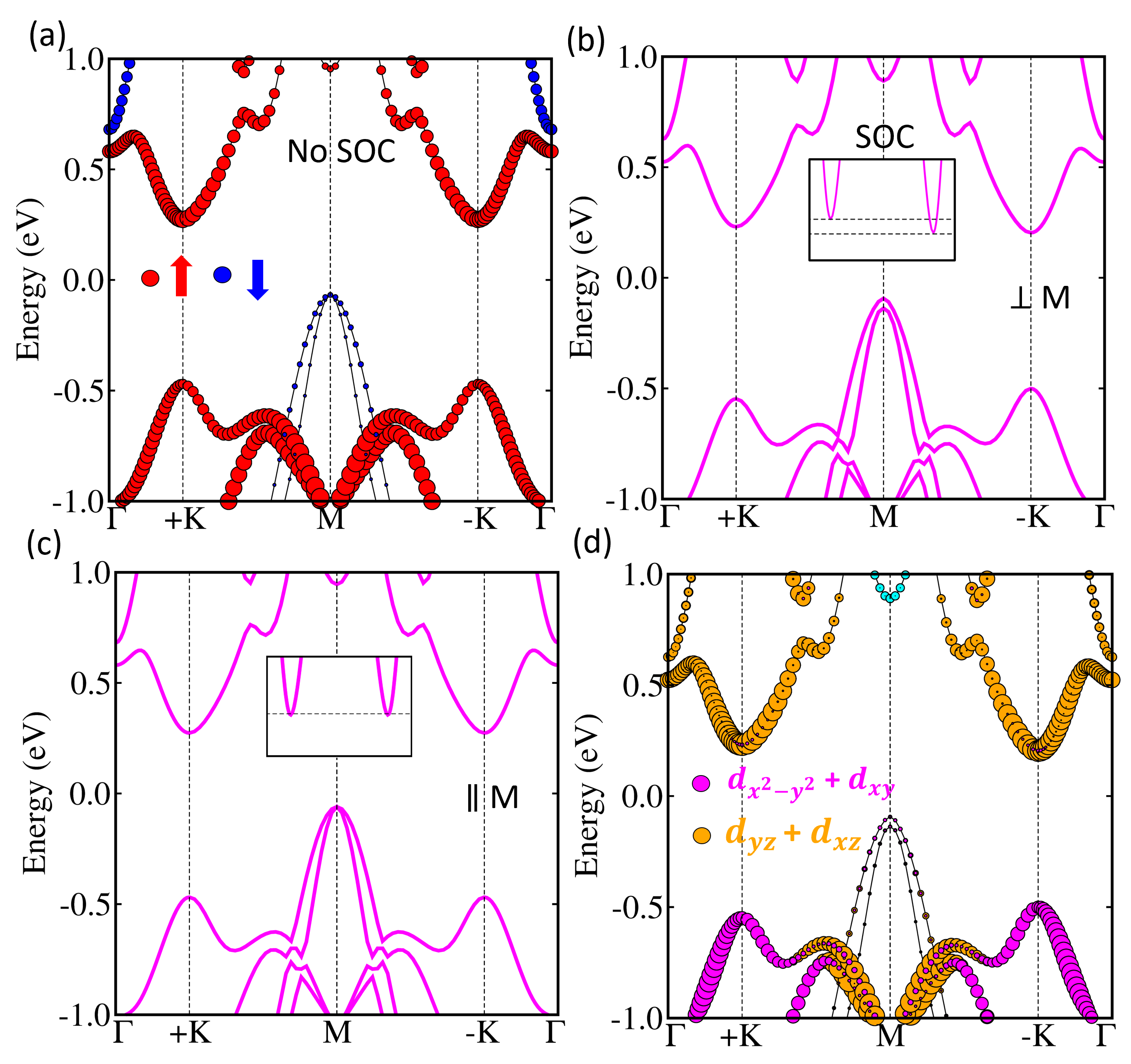}
	\caption{Band structure of Cr$_2$CSF (a) without SOC, (b) with SOC for out-of-plane magnetization (inset is showing the valley splitting in conduction band). (c) Orbital projected band structure for Cr atom. (d) Band structure for in-plane magnetization.}
	\label{pic3}
\end{figure}

We have further turned to the electronic structure calculations of Cr$_2$CSF MXene. Figure \ref{pic3}(a) presents the band structure of Cr$_2$CSF without SOC. Because of the intrinsic spin-polarization, the spin-up and spin-down channels split significantly. The bands near Fermi level are contributed by spin-down states in the valence band and spin-up states in the conduction band. As the extrema of conduction and valence bands arise from opposite spin channels, it is therefore called a bipolar FM semiconductor ~\cite{gao2022bipolar}. Cr$_2$CSF has an indirect band gap of  0.34 eV. Furthermore, the conduction band minimum (CBM) is situated at $+$K and $-$K points, giving rise to a pair of energetically equivalent valleys within the conduction band. In case of the valence band, although the valence band maximum (VBM) is located at the $\Gamma$ point, local extrema at the $+$K and $-$K points are also formed, providing monolayer Cr$_2$CSF with a pair of degenerate valleys within the valence band also. Both of these valley pairs originate from the spin-up channels. When SOC is included, the valley at $+$K shifts upwards and $-$K shifts downwards, producing a spontaneous valley polarization of 27.1 meV in the conduction band (see Fig. \ref{pic3}(b)). Whereas in the valence band, the valley at $+$K shifts downwards and $-$K shifts upwards, leading to a spontaneous valley polarization of 46.8 meV, giving a total valley polarization of 73.9 meV. However, the valleys in valence band lie far away from the Fermi level, rendering them not suitable for practical applications. Therefore, we have solely focused on the pair of valleys in the conduction band. 
Alongside, we have plotted the band structures for Cr$_2$CSX (X= O, H, Cl, Br, and I) [see Sec. V of SM]. Cr$_2$CSCl and Cr$_2$CSI show a small valley polarization of 19.8 meV and 0.8 meV, respectively, in the conduction band, whereas no significant valley polarization has been observed for Cr$_2$CSO, Cr$_2$CSI and Cr$_2$CSH. 

The spontaneous valley polarization in monolayer Cr$_2$CSF is attributed to the combined effects of time-reversal symmetry breaking, inversion symmetry breaking, and the presence of a strong SOC interaction. In addition to above, the out-of-plane magnetic anisotropy is essential for spontaneous polarization of valleys ~\cite{he2021two}. To show this, we have plotted the band structure of Cr$_2$CSF by taking the magnetization direction as in-plane and no valley polarization has been observed in this case [Fig. \ref{pic3} (c)]. To gain a deeper understanding of the influence of magnetization orientation on valley polarization, we have employed a simple model to describe the SOC term as follows ~\cite{wang1996torque,dai2008effects}:

\begin{equation}
\hat{H}_{\text{SOC}}^0 + \hat{H}_{\text{SOC}}^1 = \lambda \hat{S}_{z^\prime}\hat{L}_z
\end{equation}

\noindent Here, $\hat{H}_{\text{SOC}}^0$ represents the interaction between identical spin states, while $\hat{H}_{\text{SOC}}^1$ represents the interaction between opposite spin states. \(\hat{L_z}\) and \(\hat{S_{z^\prime}}\) represent the operators for orbital angular momentum and spin angular momentum, respectively. Further, due to the FM spin splitting, the valleys are predominantly contributed by either the spin-up or spin-down states. In Cr$_2$CSF, both the valleys of conduction band are contributed by spin-up states only and interactions between opposite spin states have a negligible effect on the valleys. Therefore, we have focused on same spin states and $\hat{H}_{\text{SOC}}^1$ has been neglected. $\hat{H}_{\text{SOC}}^0$ can be expressed as:

\begin{equation}
\hat{H}_{\text{SOC}}^0 = \lambda \hat{S}_{z^\prime} \left (\hat{L}_z \cos\theta + \frac{1}{2} \hat{L}_+ e^{-i\phi} \sin\theta + \frac{1}{2} \hat{L}_- e^{i\phi} \sin\theta\right)
\end{equation}
\noindent Here, the magnetization is out-of-plane, i.e., \(\theta\) = 0. Therefore, the above equation is simplified to:

\begin{equation}
\hat{H}_{\text{SOC}}^0 = \lambda \hat{S}_{z^\prime} \hat{L}_z 
\end{equation}

\noindent To proceed further, we have seen in Fig. \ref{pic3} (d) that the valleys in conduction band are predominantly composed of $d_{xz}$ and $d_{yz}$ orbitals of the Cr atoms. Therefore, the basis wave vectors are chosen in the following form:

\begin{equation}
|\phi_{c}^\tau\rangle = \frac{1}{\sqrt{2}} \left(|d_{xz}\rangle + i\tau|d_{yz}\rangle \right) \otimes |\uparrow\rangle
\end{equation}
where c represents the conduction band and $\tau$ = $\pm$1 corresponds to $+$K/$-$K valley. The energy level at $+$K/$-$K can be denoted by following formula:
\begin{equation}
\begin{aligned}
E_{c}^{+\mathrm{K}/-\mathrm{K}} &= \langle\phi_{c}^\tau|\hat{H}_{\text{SOC}}^0|\phi_{c}^\tau\rangle \\
& = \frac{i\tau}{2}\langle d_{xz},\uparrow|\hat{H}_{\text{SOC}}^0|d_{yz},\uparrow\rangle\\ & - \langle d_{yz},\uparrow|\hat{H}_{\text{SOC}}^0|d_{xz},\uparrow\rangle\\
& = \frac{i\tau}{2}\lambda\beta\langle d_{xz}|\hat{L}_z|d_{yz}\rangle\ -\langle d_{yz}|\hat{L}_z |d_{xz}\rangle
\end{aligned}
\end{equation}
 where $\beta$ = $\braket{\uparrow|\hat S_z|\uparrow}$, \(\hat{L_z}\)$\ket{d_{xz}}$ = $\textit{i}$$\ket{d_{yz}}$, and \(\hat{L_z}\)$\ket{d_{yz}}$ = --$\textit{i}$$\ket{d_{xz}}$. Therefore, the valley polarization in the conduction band of Cr$_2$CSF is obtained through

\begin{equation}
\begin{aligned}
\Delta E_{c} & = E_{c}^{+\mathrm{K}} - E_{c}^{-\mathrm{K}} \\
 & = i\lambda\beta\langle d_{xz}|\hat{L}_z|d_{yz}\rangle - i\lambda\beta\langle d_{yz}|\hat{L}_z|d_{xz}\rangle\\
 & = 2\lambda\beta
\end{aligned}
\end{equation}

\noindent On the other hand, if magnetic orientation is in-plane, i.e., $\theta = \frac{\pi}{2}$, we get:

\begin{equation}
\hat{H}_{\text{SOC}}^0 = \lambda \hat{S}_{z^\prime} \left ( \frac{1}{2} \hat{L}_+ e^{-i\phi}  + \frac{1}{2} \hat{L}_- e^{i\phi} \right)
\end{equation}

\noindent In this case, \textit{$\Delta$\textit{E}$_{c}$} = 0, i.e., no spontaneous valley polarization. Therefore, we can conclude that SOC and out-of-plane magnetic orientation are necessary for achieving spontaneous valley polarization.
 
Further, Berry curvature has been an invaluable tool for examining the Hall effect. Particularly in hexagonal systems where space inversion symmetry is broken, it has been observed that charge carriers within the $+$K and $-$K valleys tend to acquire a non-zero Berry curvature. Furthermore, when time reversal symmetry is also broken, an interesting characteristic emerges: a contrasting feature between valleys becomes apparent. Therefore, to understand the valley-related properties of Cr$_2$CSF we have calculated the Berry curvature using Kubo formula ~\cite{thouless1982quantized}: 
\begin{equation}
\Omega (\mathrm{K}) = - \sum_{n} \sum_{m \neq n} f_n \frac{2 Im \langle \psi_{nk} | v_x | \psi_{mk} \rangle  \langle \psi_{mk} | v_y | \psi_{nk} \rangle} {(E_n - E_m)^2}
\end{equation}
Here, \(f_n\) is the Fermi-Dirac distribution function, \(E_{n}\) represents the eigenvalue of the Bloch state \(|\psi_{nk}\rangle\) and \(\hat{v}_x\) and \(\hat{v}_y\) are the velocity operator components. 

\begin{figure}[htb]
	\includegraphics[width=0.5\textwidth]{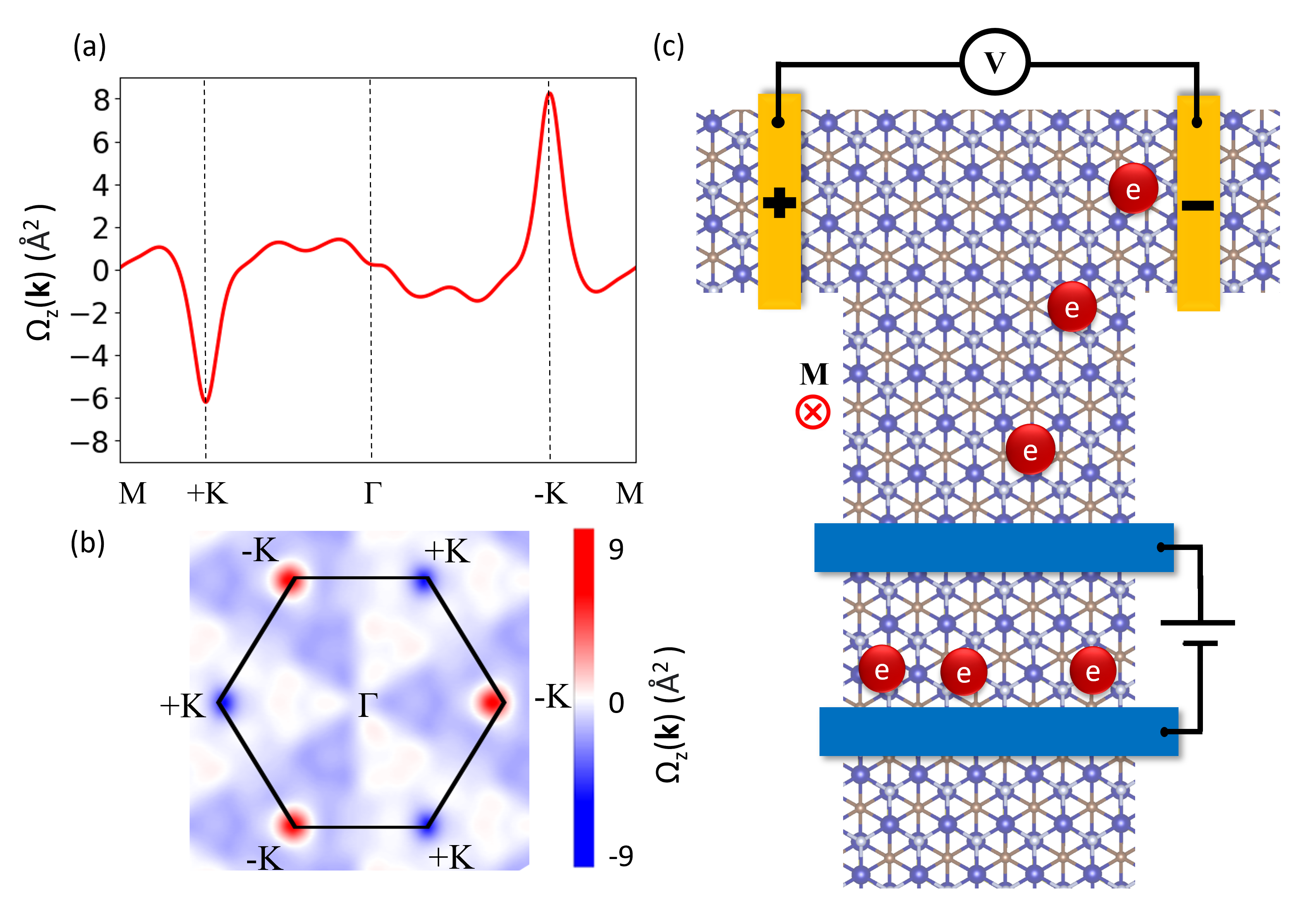}
	\caption{(a) Berry curvature of Cr$_2$CSF along the high symmetry path. (b) Contour diagram of Berry curvature. (c) Schematic of AVH in Cr$_2$CSF under the in-plane electric field and electron doping showing the movement of electrons towards right edge of the sample.}
	\label{pic4}
\end{figure}

\begin{figure}[htp]
	\includegraphics[width=0.5\textwidth]{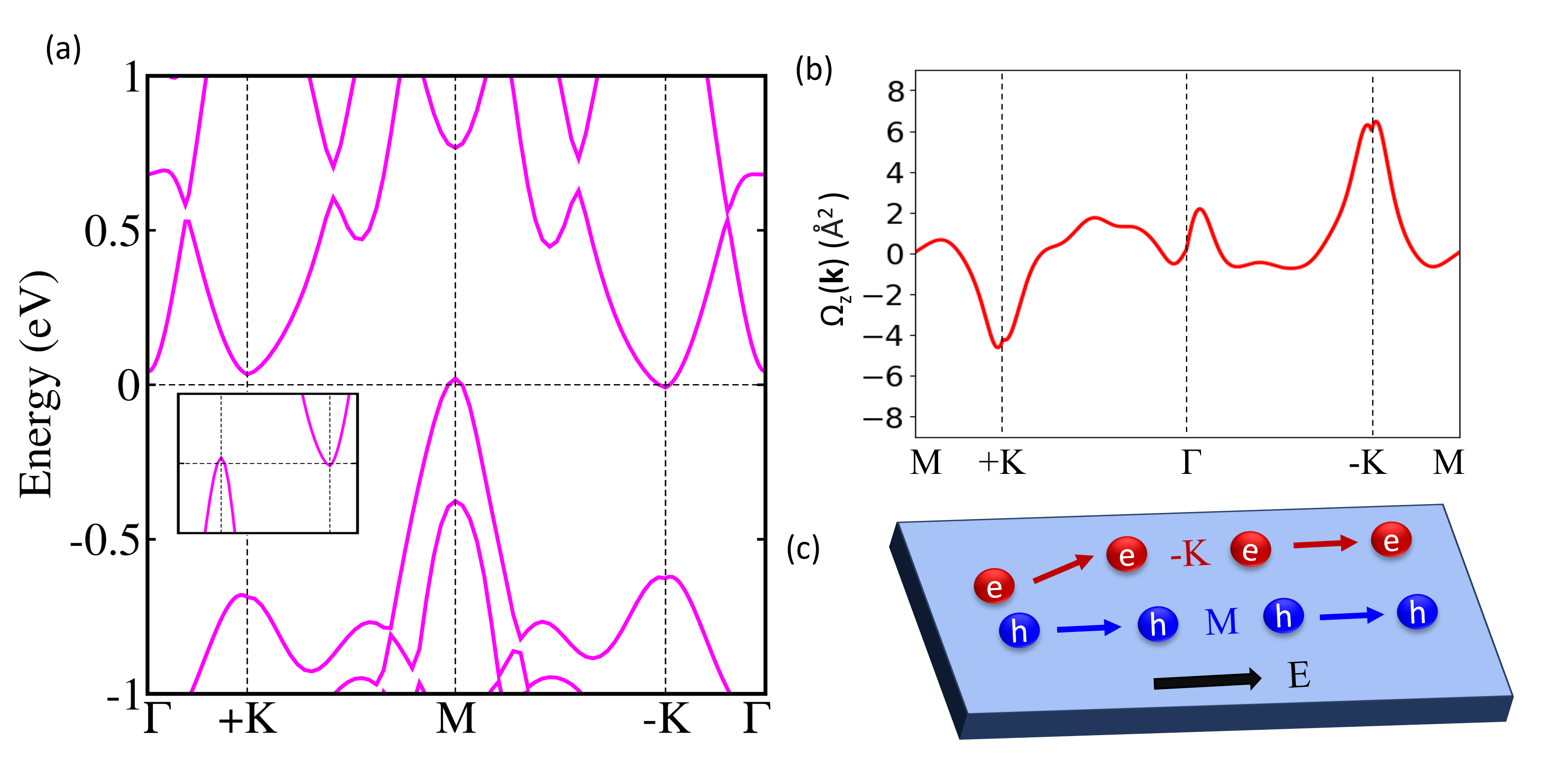}
	\caption{(a) Band structure of Cr$_2$CSF under uniaxial strain of 3.5\% showing QHVM phase. (b) Berry curvature of Cr$_2$CSF along the high symmetry path. (c) Schematic of hole-electron charge carriers separation on application of in-plane electric field.}
	\label{pic5}
\end{figure}

The Berry curvature as a curve along the path and contour map over the 2D Brillouin zone is shown in Fig. \ref{pic4}(a) and (b), respectively. At the high symmetry points $+$K and $-$K, the Berry curvature exhibits opposite signs, with absolute magnitudes of $-$6.21 and 8.25 \AA$^2$ at $+$K and $-$K, respectively. Consequently, the integral of Berry curvature around one valley differs from that around the other, leading to measurable Hall conductivity, known as the AVH effect. The anomalous Hall effect in FV materials is termed as AVH effect because an additional component of Hall conductivity is added which occurs due to the spontaneous valley polarization. AVH can be observed in Cr$_2$CSF under the effect of an in-plane electric field. Under this field, the Bloch electrons will gain an anomalous velocity, which is equal to the cross product of in-plane electric field and Berry curvature i.e., \(\upsilon \sim E \times \Omega(\mathbf{k})\) ~\cite{sheoran2023manipulation,xu2014spin,xiao2010berry}. Now, via electron doping the Fermi level can be shifted to an energy position between $-$K valley and $+$K valley. The spin-up electrons in $-$K valley will start moving with anomalous velocity and get accumulated on the right side of the sample as seen in schematic Fig. \ref{pic4}(c).

The effect of external strain has been well-known in modulation of electronic properties ~\cite{sheoran2023probing,hu2019two,lei2019broken,zang2021large,gunawan2006valley}. Consequently, we have applied uniaxial strain in \textit{y}-direction to understand the effect of strain on valley-related properties of Cr$_2$CSF. On reducing the \textit{y}-dimension in Cr$_2$CSF by 3.5\%, an intriguing QHVM has been observed. In QHVM, the CBM and VBM slightly touch the Fermi level such that both the electron and hole carriers coexist but only one type of charge carriers are valley polarized. As seen in the electronic band structure in Fig. \ref {pic5}(a), CBM touches the Fermi level at $-$K while VBM touches the Fermi level at M. The Berry curvature along the 2D Brillouin zone only occurs around $+$K and $-$K valleys with opposite signs and unequal magnitudes ($-$4.21 and 6.15 \AA$^2$) as shown in Fig. \ref{pic5}(b). On application of in-plane electric field, the electronic carriers of the $-$K valley turn towards one edge of the sample while the hole carriers of the M valley move in a straight line along the applied electric field (see Fig. \ref{pic5}(c)). Therefore, we can detect the AVH effect in Cr$_2$CSF by direct electric measurement, facilitating the practical valleytronic applications. The present research provides a tantalizing candidate for realizing and manipulating valley and spin physics.\\




\section{CONCLUSIONS}
To summarize, using first-principles calculations, we have studied the valley-dependent properties of Cr$_2$CSX MXenes. Our results reveal that monolayer Cr$_2$CSF is an indirect band gap FM semiconductor exhibiting an intrinsic out-of-plane magnetization with a large MAE value of 1.16 meV. This out-of-plane magnetization and time-reversal symmetry breaking leads to spontaneous valley polarization of 27.1 meV in the conduction band. The underlying physics of spontaneous valley polarization has also been discussed in detail based on the SOC Hamiltonian model. This spontaneous valley polarization causes observation of AVH effect under in-plane electric field and electron doping, leading to a separation of charge carriers. Additionally, we have revealed that the valley-related switchable Hall effects can be effectively engineered by the application of strain. On application of uniaxial strain of 3.5\% in Cr$_2$CSF, QHVM has been observed. The hole and electron charge carriers are separated in the presence of in-plane electric field. Existence of switchable Hall effects and presence of very high out-of plane MAE in this material makes it promising for applications in valleytronics.


\section*{ACKNOWLEDGEMENT}
	A.P. acknowledges IIT Delhi for the senior research fellowship. S.S. acknowledges CSIR, India, for the senior research	fellowship [grant no. 09/086(1432)/2019-EMR-I]. S.B. acknowledges financial support from SERB under a core research grant (grant no. CRG/2019/000647) to set up his High Performance Computing (HPC) facility ‘‘\textit{Veena}’’ at IIT Delhi for computational resources.

	

\end{document}


\begin{flushleft}	
\title{Probing switchable valley-related Hall effects in 2D magnetic MXenes. }
\author{Ankita Phutela\footnote{Ankita@physics.iitd.ac.in}, Sajjan Sheoran, Saswata Bhattacharya\footnote{saswata@physics.iitd.ac.in}} 
\affiliation{Department of Physics, Indian Institute of Technology Delhi, New Delhi 110016, India}
\maketitle
\begin{center}
	{\Large \bf Supplemental Material}\\ 
\end{center}
\begin{enumerate}[\bf I.]
	\item Crystal structures of four types of functionalization adsorption sites on Cr$_2$CSX MXenes.
	\item Energies of different configurations of Cr$_2$CSX MXenes.
	\item Energies of different configurations of Cr$_2$CSF MXenes using HSE06 functional.
	\item Crystal structures and energies of ferromagnetic and antiferromagnetic states for Cr$_2$CSF MXene.
	\item Band structures of Cr$_2$CSX (X = H, O, Cl, Br, I) MXenes.

\end{enumerate}
\vspace*{12pt}
\clearpage
\section{C\MakeLowercase{rystal structures of four types of functionalization adsorption sites on} C\MakeLowercase{r}$_2$CSX MX\MakeLowercase{enes}.}

\begin{figure}[h]
	\includegraphics[width=0.7\textwidth]{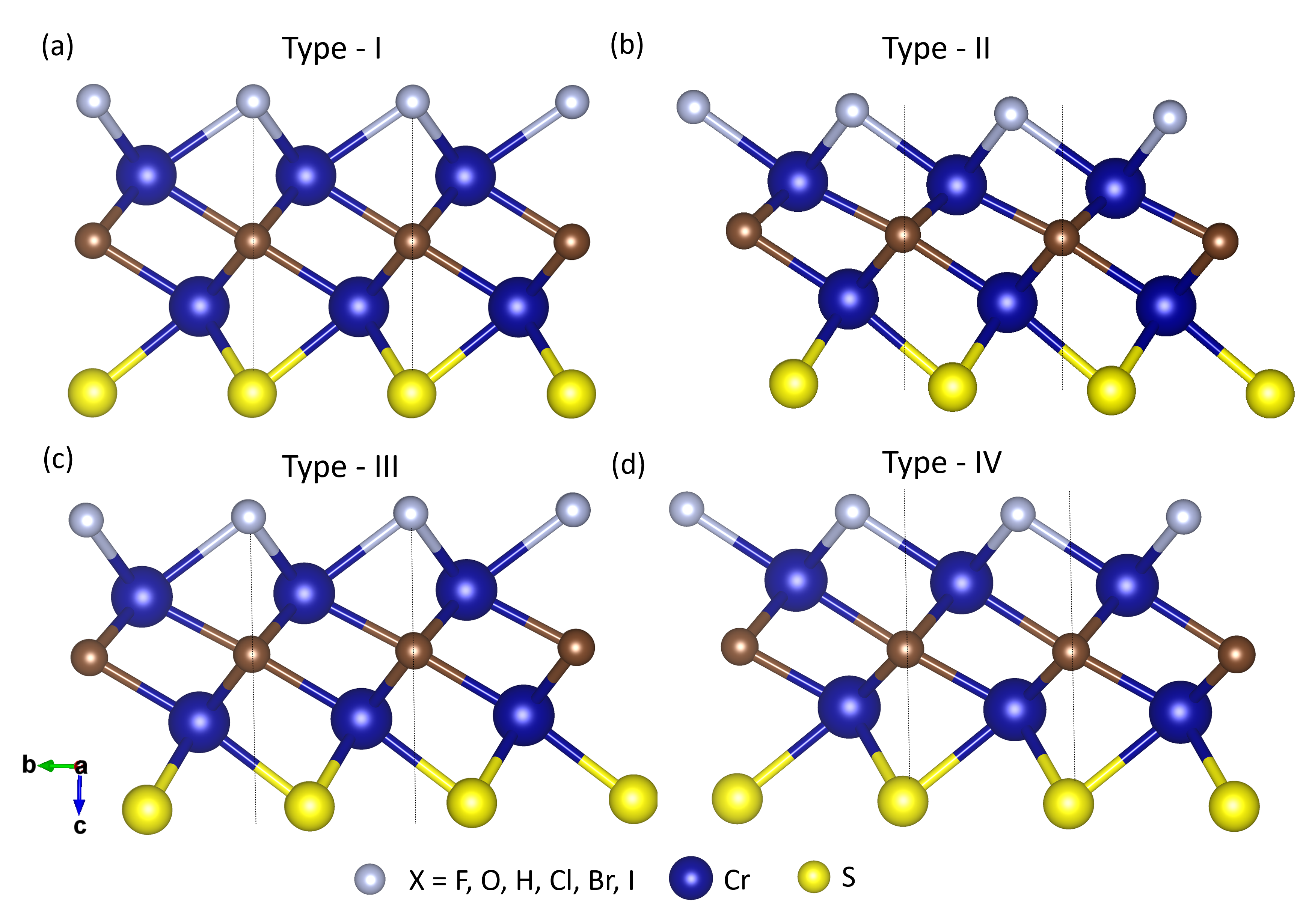}
	\caption{Crystal structure of (a) Type I, (b) Type II, (c) Type III, and (d) Type IV  Cr$_2$CSX (X = F, O, H, Cl, Br, I) MXene.}
	\label{pic3}
\end{figure}
\newpage

\section{E\MakeLowercase{nergies of different configurations of} C\MakeLowercase{r}$_2$CSX MX\MakeLowercase{enes}.}
\begin{table*}[htbp]
	\caption {Four types of functionalization adsorption sites on Cr$_2$CSX MXenes. The most stable
		type has been set up to zero for each MXene (in units of eV).}
	\begin{ruledtabular}	
		\begin{tabular}[c]{cccccccccccc} 		
			MXenes& Type I & Type II & Type III &Type IV    \\ \hline
			Cr$_2$CSH  & 0.83  & 0 & 0.63 & 0.19 \\ 
			Cr$_2$CSF  & 0.72  & 0 & 0.58 & 0.22 \\ 
			Cr$_2$CSCl  & 0.53 & 0 & 0.14 & 0.05 \\
			Cr$_2$CSBr  & 0.37  & 0 & 0.02 & 0.06 \\
			Cr$_2$CSI  & 0.46  & 0 & 0.25 & 0.26 \\ 
			Cr$_2$CSO  & 1.13 & 0 & 0.87 & 0.05 \\
		\end{tabular}
	\end{ruledtabular}
	\label{T1}
\end{table*}

\section{E\MakeLowercase{nergies of different configurations of} C\MakeLowercase{r}$_2$CSF MX\MakeLowercase{enes using} HSE06 \MakeLowercase{functional}.}
\begin{table*}[htbp]
	\caption {Energies of different types of configurations of Cr$_2$CSF MXene using HSE06 functional. The most stable	type has been set up to zero (in units of eV).}
	\begin{ruledtabular}	
		\begin{tabular}[c]{cccccccccccc} 		
			MXenes& Type I & Type II & Type III &Type IV    \\ \hline
			Cr$_2$CSF  & 1.31  & 0 & 1.61 & 0.67 \\ 
			
		\end{tabular}
	\end{ruledtabular}
	\label{T2}
\end{table*}
\newpage
\section{C\MakeLowercase{rystal structures and energies of ferromagnetic and antiferromagnetic states for} C\MakeLowercase{r}$_2$CSF MX\MakeLowercase{ene}.}
\begin{figure*}[htp]
	\includegraphics[width=0.7\textwidth]{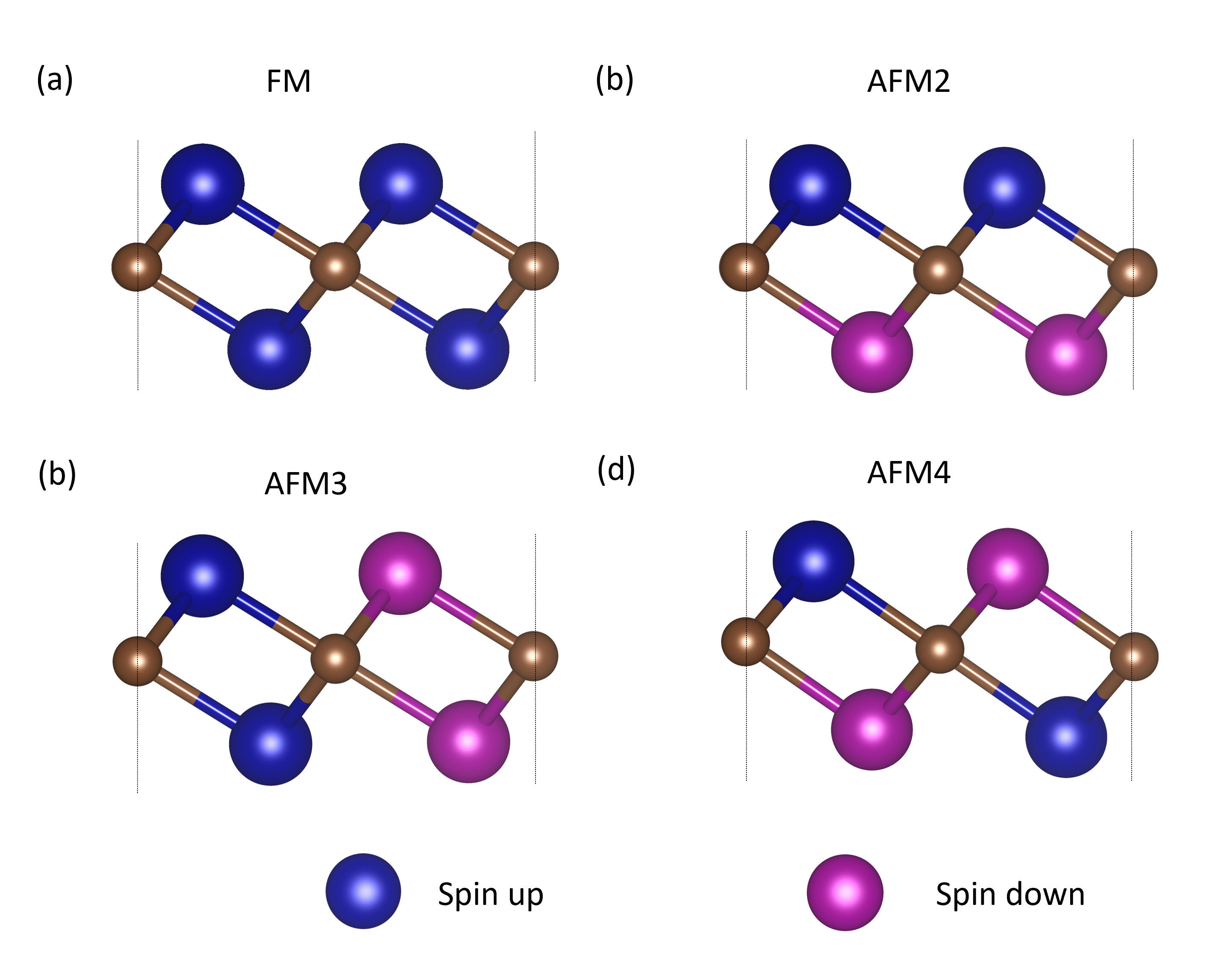}
	\caption{Crystal structure of (a) one FM and (b), (c), (d) three AFM states of Cr$_2$C MXene.}
	\label{p1}
\end{figure*}

\begin{table*}[htbp]
	\caption {Four types of magnetic configuartions of Cr$_2$CSF MXene. The most stable
		type has been set to zero (in units of eV).}
	\begin{ruledtabular}	
		\begin{tabular}[c]{cccccccccccc} 		
			Cr$_2$CSF MXene& PBE+U & HSE06    \\ \hline
			FM  & 0  & 0  \\ 
			AFM1  & 1.64  & 0.9  \\ 
			AFM2  & 1.53 & 1.55  \\
			AFM3  & 1.83  & 0.95  \\
		\end{tabular}
	\end{ruledtabular}
	\label{T3}
\end{table*}

\newpage

\section{B\MakeLowercase{and structures of} C\MakeLowercase{r}$_2$CSX (X = H, O, C\MakeLowercase{l}, B\MakeLowercase{r}, I) MX\MakeLowercase{enes}.}
\begin{figure*}[htp]
	\includegraphics[width=0.9\textwidth]{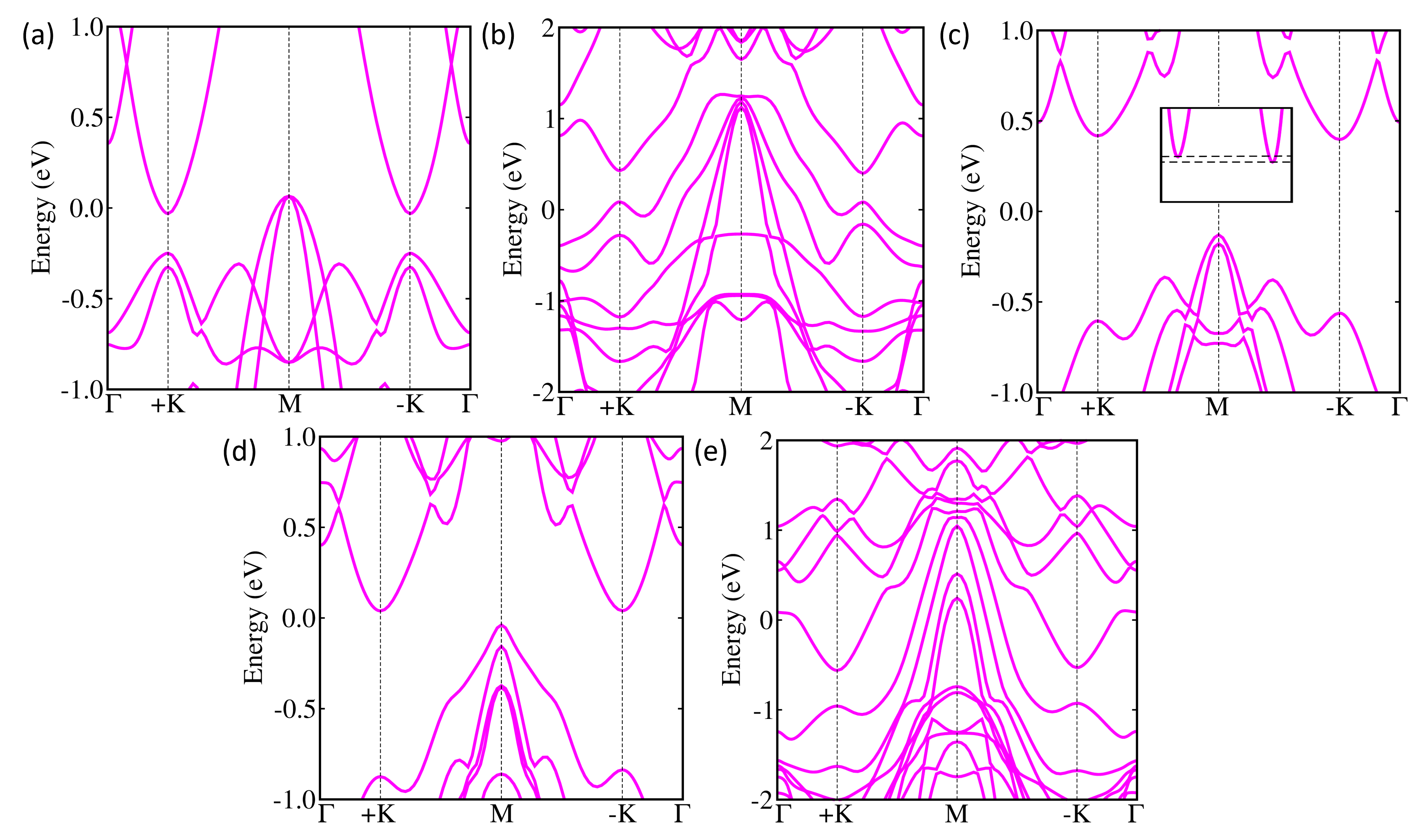}
	\caption{Band structure of (a) Cr$_2$CSH, (b) Cr$_2$CSO, (c) Cr$_2$CSCl, (d) Cr$_2$CSBr, (e) Cr$_2$CSI MXene in presence of SOC.}
	\label{p2}
\end{figure*}
\newpage


\end{flushleft}